\documentclass[12pt]{article}
\usepackage{amssymb}
\usepackage{amsmath}
\usepackage{hyperref}
\usepackage{graphicx}
\usepackage{placeins}
\usepackage[font=small,labelfont=bf]{caption}
\usepackage{bm}
\usepackage{latexsym}
\begin{document}
\title{\bf    Non-Abelian Exponential Yang-Mills AdS Black Brane and Transport Coefficients}
\author{  Mehdi Sadeghi\thanks{Corresponding author: Email:  mehdi.sadeghi@abru.ac.ir}  and  Faramaz Rahmani\thanks{Email:  faramarz.rahmani@abru.ac.ir}\hspace{2mm}\\
{\small {\em Department of Physics, Faculty of Basic Sciences,}}\\
        {\small {\em Ayatollah Boroujerdi University, Boroujerd, Iran}}
       }
\date{\today}
\maketitle 

\abstract{ In this paper, AdS black brane solution of Einstein-Hilbert gravity with non-abelian exponential guage theory of Yang-Mills type is introduced. DC conductivity and the ratio of shear viscosity to entropy density as two important transport coefficients are calculated by using of Kubo formula in the context of AdS/CFT duality. Our results recover the Yang-Mills model in $q\to \infty$ limit.}\\

%\pacs{11.10.Jj, 11.10.Wx, 11.15.Pg, 11.25.Tq}


%--------------------------------------------------------------------------
\section{Introduction} \label{intro}

Quantum chromodynamics (QCD) as a non-abelian gauge theory with $SU(3)$ symmetry group describes the strong interaction between quarks mediated by gluons. The strong interaction can be described through the Yang-Mills theory with the Lagrangian density, ${\bf{Tr}}( F_{\mu \nu }^{(a)} F^{(a)\, \, \mu \nu })$. The Yang-Mills theory is a nonlinear theory and it suffers from problems such as singularity on the point like charge and infinite self-energy \cite{Delphenich:2003yw,Delphenich:2006ec,EslamPanah:2021xaf}. 
The Lagrangian of nonlinear Yang-Mills field has been considered in various theories of gravity. The generalization of Yang-Mills black holes plays a crucial role since most of physical systems are intrinsically nonlinear in the nature which leads to revelation of some loop corrections.\par 
The Einstein-Yang-Mills AdS black brane \cite{Sadeghi:2014uqf} is a solution of Einstein-Yang-Mills gravity with the flat topology of event horizon. The generalization of this solution is studied in various gravity theories like Gauss-Bonnet massive gravity\cite{Sadeghi:2015vaa} and higher derivative massive gravity\cite{Parvizi:2017boc} to explain inflation and current acceleration of the Universe \cite{Camara:2004ap}-\cite{Garcia-Salcedo:2013cxa} and some aspects of quantum gravity.\par 
Logarithmic gauge theory \cite{Gaete:2013dta},\cite{Dehghani:2019xjc},\cite{Sadeghi:2022mog}, arcsin-electrodynamics \cite{Kruglov_2015},\cite{Sadeghi:2023rsn}, Born-Infeld theory \cite{Born1,Born2} have been introduced as nonlinear theories (NED) to solve some of problems in Yang-Mills theory. The NED fields give more information in higher magnetized neutron stars and pulsars \cite{Bialynicka-Birula}\cite{H. J. Mosquera}. These theories remove both of big-bang and black hole singularities \cite{Corda:2009xd}-\cite{AyonBeato:1999ec} by modifying spacetime geometry.\par 
There are three kinds of Born-Infeld (BI) actions, Born-Infeld nonlinear electromagnetic (BINEF), logarithmic form of nonlinear electromagnetic field (LNEF) and exponential form of nonlinear electromagnetic field (ENEF)\cite{Hendi:2013dwa}.\\
\begin{equation}
L(\mathcal{F})= \left\{ \begin {array} {cc} 4 q^2 \bigg(1-\sqrt{1+\frac{\mathcal{F}}{2 q^2}}\bigg)  & \text{BINEF} \\q^2 \bigg(exp({-\frac{\mathcal{F}}{q^2}})-1\bigg)  & \text{ENEF} \\ -8 q^2 \ln\bigg(1+\frac{\mathcal{F}}{8q^2}\bigg) & \text{LNEF}. \\   \end {array} \right.
\end{equation}
These Lagrangian functions tends to the Yang-Mills Lagrangian functions when, $q \to \infty$.
A good question that can be asked is, what is the need to introduce these nonlinear models?. Why is the specific nonlinear YM action (ENEF) chosen among other options?. To answer these questions, we consider that above Lagrangian functions can be expanded as following general form,
\begin{equation}\label{L1}
	\mathcal{L}=-\mathcal{F}+\alpha \mathcal{F}^2+ \beta \mathcal{F}^3+ \gamma \mathcal{F}^4+...\,\,\,.
\end{equation}
Here, $\alpha$ , $\beta$ and $\gamma$ are the general constant coefficients and $\mathcal{F}= F_{\mu \nu } F^{\mu \nu }$. For each of above functions, the expansion coefficients are different. So, we have: 
\begin{equation}
	L(\mathcal{F})= \left\{ \begin {array} {cc} 4 q^2 \bigg(1-\sqrt{1+\frac{\mathcal{F}}{2 q^2}}\bigg)=-\mathcal{F}+\frac{\mathcal{F}^2}{8 q^2}-\frac{\mathcal{F}^3}{32 q^4}+\frac{5 \mathcal{F}^4}{512 q^6}-\frac{7 \mathcal{F}^5}{2048 q^8}+...  & \text{BINEF} \\q^2 \bigg(exp({-\frac{\mathcal{F}}{q^2}})-1\bigg)=-\mathcal{F}+\frac{\mathcal{F}^2}{2 q^2}-\frac{\mathcal{F}^3}{6 q^4}+\frac{ \mathcal{F}^4}{24 q^6}-\frac{\mathcal{F}^5}{120 q^{8}}+...  & \text{ENEF} \\ -8 q^2 \ln\bigg(1+\frac{\mathcal{F}}{8q^2}\bigg)=-\mathcal{F}+\frac{\mathcal{F}^2}{16 q^2}-\frac{\mathcal{F}^3}{192 q^4}+\frac{\mathcal{F}^4}{2048 q^6}-\frac{ \mathcal{F}^5}{20480 q^8}+... & \text{LNEF}. \\   \end {array} \right.
\end{equation}
The general behavior of these functions may be the same, but they differ in details. Different expansion coefficients can help us to describe the cosmic inflation periods and match the observational results \cite{Hendi:2013dwa},\cite{Hendi:2012zz}. Therefore, examining each of the functions allows us to test our luck in finding an explanatory optimal higher derivative YM models that match the observations.\par 
The color DC conductivity bound of Einstein–Born–Infeld AdS black brane was studied in \cite{Sadeghi:2021qou} and the same method for logarithmic form of Born-Infeld (BI) action was considered in \cite{Sadeghi:2022mog}. We consider the exponential form of nonlinear Yang-Mills in this paper.\par 
Fluid-gravity duality  \cite{Son}-\cite{J.Bhattacharya2014} is a version of AdS/CFT duality\cite{Maldacena:1997re} for long wavelength limit. The boundary theory behaves as a fluid effectively and hydrodynamics equations which are conserved charges up to first order of derivative expansion are as follows,  
%We want to study the field theory dual of the logarithmic gauge theory by AdS/CFT duality \cite{Maldacena:1997re}. Therefore, we should calculate the transport coefficients like conductivity and shear viscosity to entropy density.
%Hydrodynamics is an effective theory for describing near-equilibrium phenomena at late
%time and large distance \cite{Landau,Son,Policastro2001}. Conservation equations together with constitutive relations in a gradient expansion around equilibrium is constructed hydrodynamics equations. These equations up to first order of derivative expansion are as follows, \\
\begin{align}
& \nabla _{\mu } J^{\mu } =0\,\,\,\,\,,\,\,\nabla _{\mu } T^{\mu \nu} =0,\\
& J^{\mu } =n \, u^{\mu }-\sigma T P^{\mu \nu }\partial_{\nu}(\frac{\mu}{T}),\nonumber\\
& T^{\mu \nu } =(\rho +p)u^{\mu } u^{\nu } +pg^{\mu \nu } -\sigma ^{\mu \nu },\nonumber\\
&\sigma ^{\mu \nu } = {P^{\mu \alpha } P^{\nu \beta } } [\eta(\nabla _{\alpha } u_{\beta } +\nabla _{\beta } u_{\alpha })+ (\zeta-\frac{2}{3}\eta) g_{\alpha \beta } \nabla .u],\nonumber\\& P^{\mu \nu }=g^{\mu \nu}+u^{\mu}u^{\nu}, \nonumber
\end{align}
 where $n$, $\eta$, $\zeta $, $\sigma ^{\mu \nu }$ , $\sigma$ and $P^{\mu \nu }$ are charge density, shear viscosity, bulk viscosity, shear tensor, conductivity and projection operator, respectively \cite{Kovtun2012}. $\eta$, $\zeta $ and $\sigma $ are known as transport coefficients.\\
 Green-Kubo formula relates transport coefficients to two-point function \cite{Son} as follows,
\begin{equation} \label{kubo}
	\sigma^{(ab)} _{ij}(k_{\mu})=-\mathop{\lim }\limits_{\omega \to 0} \frac{1}{\omega } \Im G^{(ab)}_{ij}(k_{\mu}).
\end{equation} 
Where lower and upper indices refer to the spatial directions and gauge group, respectively. 
%The fluid-gravity correspondence as a subclass of gauge-gravity duality states that
%black holes in AdS space-time are dual to stationary solutions of the equations of relativistic hydrodynamics \cite{Son}-\cite{J.Bhattacharya2014}.  Conductivity is calculated by Green-Kubo formula and Witten prescription.\\ 
%where $a,b$ and $i,j$ indices refer to  $SU(2)$ group symmetry and translational symmetry in spatial directions of $x,y$, respectively. 
%There is no difference between upper and lower indices on the conformal field theory (CFT) side because the metric of the boundary theory is flat spacetime with Minkowski metric. 
There is a universal bound for DC conductivity as $\sigma \geq \frac{1}{e^2}$ \footnote{We set $e=1$, so $\sigma \geq 1$ } where $e$ is the charge of the gauge field. This bound is also saturated in graphene\cite{Ziegler}, but it is violated for massive gravity \cite{Baggioli:2016oqk}, models with abelian case \cite{Baggioli:2016oju} and non-abelian Born-Infeld theory\cite{Sadeghi:2021qou}.\\
 The electrical conductivity of the QGP using the AdS/CFT duality can be expressed as follows\cite{Caron-Huot:2006pee}:
\begin{equation}
	\frac{\sigma}{T}=\frac{e^2 N_c^2}{16 \pi}=0.017.
\end{equation}
This result is consistent with lattice data and the Gribov prescription\cite{Madni:2024ubw}.\\
Another important quantity in Quak-Gloun-Plasma (QGP) or systems with strong coupling is the ratio of shear viscosity to entropy density i.e. $\frac{\eta }{s} \ge \frac{\hbar }{4\, \pi \, k_{B} }$, called the Kovtun-Son-Starinet (KSS) bound \cite{Kovtun:2004de}. The value of $\frac{\eta }{s}$ for QGP in RHIC experiments is approximately $\frac{1}{4\pi}$, indicating that QGP exhibits fluid-like behavior with a remarkably low $\frac{\eta }{s}$. The study carried out in \cite{Baggioli:2023yvc} focuses on the ratio of $\frac{\eta }{s}$ in a non-abelian holographic p-wave superfluid model demonstrates that rotational symmetry breaking does not lead to a violation of the KSS bound.\\
For a pure $SU(3)$ gauge theory at finite temperature, the ratio  $\frac{\eta }{s}$ is given by lattice Monte-Carlo simulation as follows\cite{Meyer:2007ic},
\begin{equation}
\frac{\eta }{s}=\begin{cases}
	0.134(33) \quad (T=1.65T_c)\\
	0.102(56) \quad (T=1.24T_c).
\end{cases}
\end{equation}
In graphene,  $\frac{\eta }{s}$ has been calculated using quantum kinetic theory, and this value is very close to $\frac{1}{4\pi}$\cite{Mueller}. In other words, graphene behaves similarly to strongly-coupled materials such as QGP. The strength of interactions among excitations in a quantum fluid can be assessed by examining the value of $\frac{\eta }{s}$.\\
The KSS bound is saturated for Einstein-Hilbert gravity but it is violated for Horndeski theory \cite{Feng:2015oea}, massive gravity for $c_i<0$ \cite{Sadeghi:2020lfe}, scalar-tensor gravity\cite{Bravo-Gaete:2020lzs}-\cite{Bravo-Gaete:2022lno}, anisotropic black brane \cite{Mamo:2012sy}, Gauss–Bonnet massive gravity\cite{Sadeghi:2015vaa}, deformed $AdS_4$–Reissner–Nordström \cite{Ferreira-Martins:2019wym}, AdS$_5$-Schwarzschild deformed black branes\cite{Ferreira-Martins:2019svk}, higher curvature massive gravity\cite{Parvizi:2017boc} and massive gravity \cite{Hartnoll:2016tri}. The quantity $\frac{\eta }{s}$ is proportional to the inverse of the square of the field theory coupling i.e. $\frac{\eta }{s} \sim \frac{1 }{\lambda^2}$. \footnote{$\lambda$\ is the coupling of field theory.}\\
In condensed matter systems like graphene, the ratio of shear viscosity to entropy density and the DC conductivity are also crucial parameters which influence the transport properties of the material. Graphene is known for its exceptional electrical and thermal conductivity, as well as its strong mechanical properties. Studying the transport properties of graphene using theoretical models and experimental measurements can provide valuable insights into its unique electronic structure, thermal behavior, and potential applications in nanoelectronics and optoelectronics.\\
In this paper, we want to study the DC conductivity and $\frac{\eta}{s}$ for non-abelian exponential Yang-Mills AdS black brane to describe the field theory dual of this model.
%--------------------------------------------------------------------------
\section{  Non-Abelian Exponential Yang-Mills AdS Black Brane}
\label{sec2}

\indent The 4-dimensional action of non-abelian  exponential gauge theory with negative cosmological constant is  as follows\cite{Hendi:2013dwa},
\begin{eqnarray}\label{action}
S=\int d^{4}  x\sqrt{-g} \bigg[R-2\Lambda -q^2\big(
e^{-\frac{\mathcal{F}}{q^2}}-1\big)\bigg],
\end{eqnarray}
where, $R$ is the Ricci scalar, $\Lambda=-\frac{3}{l^2}$ is the cosmological constant,  $l$ is the AdS radius, $\mathcal{F}={\bf{Tr}}( F_{\mu \nu }^{(a)} F^{(a)\, \, \mu \nu })$ is the Yang-Mills invariant and $q$ is the nonlinear coupling constant. The trace element stands for ${\bf{Tr}}(.)=\sum_{i=1}^{3}(.).$ \\
$F_{\mu \nu }^{(a)}$ is the $SU(2) $ Yang-Mills  field strength tensor,
\begin{align} \label{YM}
F_{\mu \nu } =\partial _{\mu } A_{\nu } -\partial _{\nu } A_{\mu } -i[A_{\mu }, A_{\nu }],
\end{align}
in which the gauge coupling constant is taken equal to one,  $A_{\nu }$'s are the $SU(2)$ gauge group Yang-Mills potentials. When $q \to \infty $ the exponential term gets transformed into standard linear non-abelian Yang-Mills term \cite{Shepherd:2015dse}.
Einstein and Yang-Mills equations are obtained through the variation of the action (\ref{action}) with respect to  $g_{\mu \nu } $ and $A^{(a)}_{\nu}$ respectively. The results of variations are as follows,
\begin{equation}\label{EH}
R_{\mu \nu }-  \tfrac{1}{2}  g_{\mu \nu }R+\Lambda g_{\mu \nu }+\tfrac{1}{2}  g_{\mu \nu }q^2 \left(e^{-\frac{F_{\rho \beta }^{(a)} F^{(a)\, \, \rho \beta }}{q^2}}-1\right)-2 F^{(a)}_{\mu  \gamma  } F^{(a)}_{\nu}\,^{\gamma}   e^{-\frac{F_{\rho \beta }^{(a)} F^{(a)\, \, \rho \beta }}{q^2}} =0,
\end{equation}
and
\begin{equation}\label{EOM2}
\nabla^{\mu}\bigg(e^{-\frac{F_{\rho \beta }^{(a)} F^{(a)\, \, \rho \beta }}{q^2}} F_{\mu \nu }^{(a)}\bigg)=0.
\end{equation}
The covariant derivative is defined as,
\begin{equation}
	\nabla_{\mu} =I  \partial_{\mu}-i e \sigma^{(a)} A^{(a)}_{\mu},
\end{equation}
here, $I$ represents the identity matrix and $\sigma^{(a)}$ represents the generators of $SU(2)$.\\
We want to introduce asymptotically AdS black brane solution in four dimensions in our model. So we consider the following ansatz,
\begin{equation}\label{metric1}
ds^{2} =-f(r)dt^{2} +\frac{dr^{2} }{f(r)} +\frac{r^2}{l^2}(dx^2+dy^2),
\end{equation}
where $f(r)$ is the blackening function which will be determined by solving Eq.(\ref{EH}).
We consider the guage field ansatz as follows,
\begin{equation}\label{background}
{\bf{A}}^{(a)} =\frac{i}{\sqrt{2}}h(r)dt\begin{pmatrix}1 & 0 \\ 0 & -1\end{pmatrix}.
\end{equation}
Here, $\sigma_3= \begin{pmatrix}1 & 0 \\ 0 & -1\end{pmatrix}$ is the generator of the $SU(2)$ symmetric group of the gauge field \cite{Shepherd:2015dse}.
The non-zero component of Eq.(\ref{EOM2}) is as follows,
\begin{equation}
2 q^2 h'(r)+r \Big(q^2+ 4h'(r)^2\Big) h''(r)=0.
\end{equation}
So, $h(r)$ is found as
\begin{equation}\label{h}
h(r)=\mu + \int^{r}  \frac{1}{2} q \sqrt{W[\frac{4 e^{\frac{2 C_1}{q^2}}}{q^2 u^4}]} du,
\end{equation}
where, $\mu$ is the constant of integration and is called chemical potential of quantum field theory that locates on the boundary of AdS spacetime in AdS/CFT duality. By applying regularity condition on the event horizon i.e. $A_t(r_h) = 0$ \cite{Dey:2015poa}, we get,
%\begin{equation}
%\mu=-\frac{r_h}{12 \lambda}\Bigg(\sqrt{8 \lambda+e^{2 c_1} r_h^4}-e^{ c_1} r_h^2\Bigg)+\frac{2 i \sqrt[4]{2} }{3
%	\sqrt{i e^{c_1}} } F\left[i \sinh ^{-1}\left(\frac{\sqrt{\frac{i e^{c_1}}{\sqrt{\lambda}}} r_h}{2^{3/4}}\right),-1\right]
%\end{equation}
\begin{equation}
\mu=-\int^{r_h}  \frac{1}{2} q \sqrt{W[\frac{4 e^{\frac{2 C_1}{q^2}}}{q^2 u^4}]} du.
\end{equation}
Here $W$ is the Lambert function or omega function or product logarithm and $C_1$ is an integration constant which is related to the Yang-Mills charge.\\
The value of gauge field on the boundary is equal to chemical potential of the system\cite{Dey:2015poa},
\begin{equation}
	\lim_{r \to \infty} A_t(r)=\mu.
\end{equation}
The non-zero components of   $F_{\mu \nu}^{(a)}$ are as follows,
\begin{equation}
	F_{tr}^{(a)} =-F_{rt}^{(a)}=h'(r)\begin{pmatrix}1 & 0 \\ 0 & -1\end{pmatrix}=-\frac{1}{2 \sqrt{2}} q \sqrt{W\left(\frac{4 e^{\frac{2 C_1}{q^2}}}{q^2 r^4}\right)}\begin{pmatrix}1 & 0 \\ 0 & -1\end{pmatrix}.
\end{equation}
Therefore, the invariant scalar $\mathcal{F}_{YM}={\bf{Tr}}( F_{\mu \nu }^{(a)} F^{(a)\, \, \mu \nu })$ of gauge field is,
\begin{equation}
	\mathcal{F}_{YM}={\bf{Tr}}( F_{\mu \nu }^{(a)} F^{(a)\, \, \mu \nu }) = F_{rt}^{(a)} F^{ rt (a)} =-\frac{1}{4} q^2 W\left(\frac{4 e^{\frac{2 C_1}{q^2}}}{q^2 r^4}\right).
\end{equation}
Our results decrease to non-abelian Yang-Mills solution when, $q=0$ .\\
The Bianchi identity is also satisfied,\\
\begin{equation}
\nabla_{\alpha }F^{(a)}_{\mu \nu}+\nabla_{\nu }F^{(a)}_{\alpha \mu}+\nabla_{\mu }F^{(a)}_{ \nu \alpha}=0.
\end{equation}
By considering $xx$ component of  Eq. (\ref{EH}), we have:
\begin{equation}\label{eom}
rf''(r)+2f'(r)+2\Lambda +q^2 r (e^{-\frac{2 h'(r)^2}{q^2}}-1) =0.
\end{equation}
Now, $f(r)$ is obtained by solving Eq.(\ref{eom}) which is given by
\begin{equation}\label{sol}
f(r)=\kappa-\frac{2m}{r}+\frac{ r^2}{l^2}+q^2\int^{r}\frac{du_1}{u_1^2}  \int^{u_1} u^2 \left(1 - e^{-\frac{2 h'(u)^2}{q^2}} \right) du,
\end{equation}
where $m$ is an integration constant. Notice $r$ is the radial coordinate that put us from bulk to the boundary.\\
We note that the topology of event horizon in our model is flat so the extrinsic curvature is zero, $\kappa=0$. \par
In string theory, a BPS state refers to a special type of state that preserves a fraction of the supersymmetry of the theory. BPS stands for Bogomolnyi-Prasad-Sommerfield, named after the physicists who first studied these states. BPS states are typically stable and carry certain properties which make them of particular interest in string theory, such as having a mass that is determined by their charge. Moreover, the charge and mass are equal in the BPS state. These states play a crucial role in understanding the dynamics of string theory and have important implications for the study of black holes, particle physics, and the nature of spacetime.\\
The stable D-brane is a BPS state in string theory. Our model has the higher orders of field strength so it is similar to the solution of D-brane of string theory. The D-branes in string theory are stable solutions so the solution of our model also is stable. Due to the charge conservation in our model, the stability of our black brane solution is maintained.\\ 
It is not impossible to obtain a general analytical solution for $f(r)$ but we can explore a little about the solution of the equation. If we consider the limit $q\longrightarrow \infty$, we expect to get the Yang-Mills metric function $f(r)$. In this regard, we use the Taylor expansion of the Lambert function which for small values of its argument can be approximated as $W(x)\sim x$. Now, relation (\ref{sol}) takes the form
\begin{equation}\label{f1}
	f= \kappa -\frac{2 m}{r^{2}}+\frac{r^{2}}{l^{2}}+\frac{q^{2}+2 C_1}{q^{2} r^{2}}+\frac{\alpha}{r}+\beta ,
\end{equation}
where, $\alpha$ and $\beta$ are integration constants. In order to get the Yang-Mills solution when $q\longrightarrow \infty$, we should take $\alpha=\beta=0$ and $\frac{q^2+2C_1}{q^2}=Q^2$. So, we conclude that $\frac{2 C_1}{q^2}=Q^2-1$ which shows that the integration constant $C_1$ is really related to the Yang-Mills charge.\par
\begin{figure}[h!]
	\centering
	\includegraphics[width=6cm]{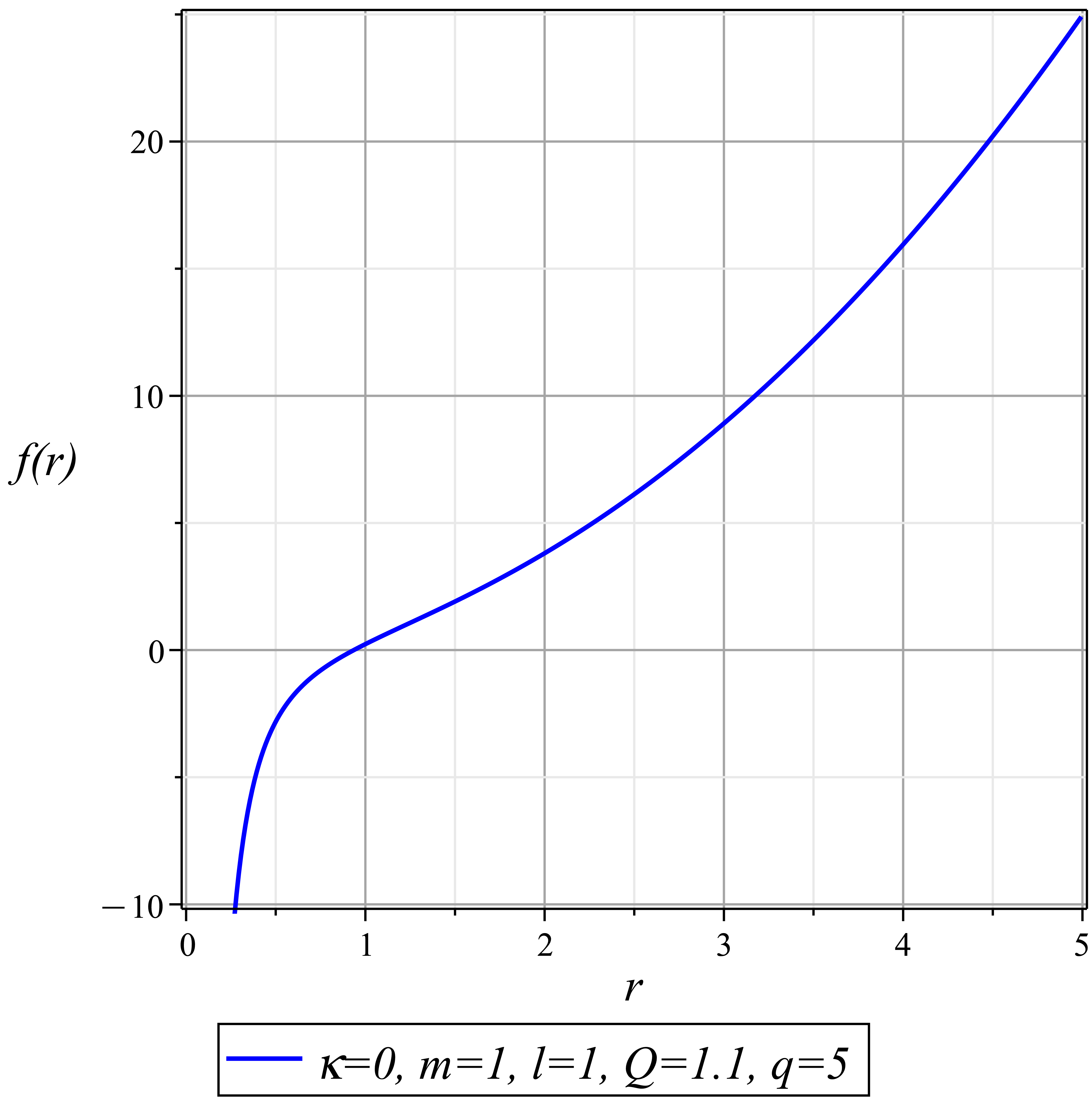}
	\caption{ An approximated behavior of $f(r)$ for the values $Q=1.1, l=1, \kappa=0, q=5$ and $m=1$. For these values the the behavior of $f(r)$ is more reliable for $r>0.6665$.  \label{fig:ff}}
\end{figure}

 
As we mentioned above, a general solution will not be achieved. But, for small arguments of the Lambert function which does not necessarily indicate the very large values of the $q$, we can use the approximation $W(x)\sim x$. For example we can assume that the problem is investigated for large values of the horizon for which the condition
\begin{equation}\label{ratio}
0< \frac{4e^{\frac{2 C_1}{q^2}}}{q^2r^4}<1,
\end{equation}
 is satisfied. However, the argument value is not very close  to zero which can be comparable with the situation in which $q\longrightarrow \infty$ and the Yang-Mills solution is obtained again. In other words, during the calculations we do not approximate expression $e^{\frac{2C_1}{q^2}}=e^{Q^2-1}$ in the argument of the Lambert function and we keep its current form. 
 According to the above approximation, we expect that $f(r)$ shows a more accurate behavior when, 
\begin{equation}\label{con}
	\displaystyle r >\sqrt{2}\, \sqrt{\frac{\sqrt{{\mathrm e}^{Q^{2}-1}}}{q}}.
\end{equation}
 After these considerations, we obtain an approximated solution which represents acceptable behavior of $f(r)$ to some extent. The result is as bellow,
\begin{equation}\label{f2}
	f = \kappa -\frac{2 m}{r^{2}}+\frac{r^{2}}{l^{2}}+\frac{{\mathrm e}^{Q^{2}-1}}{r^{2}},
\end{equation}
We see that despite the approximations we have considered, this solution is different from the Yang-Mills solution. When we take $Q=0$, the solution is not the AdS-Schwarzschild solution. 
The relation between the system parameters and the single horizon of this theory is obtained through the relation $f(r_h)=0$ which gives
\begin{equation}
	m=\frac{\kappa  r_h^{2}}{2}+\frac{r_h^{4}}{2 l^{2}}+\frac{{\mathrm e}^{Q^{2}-1}}{2}.
\end{equation}
For a black brane we must take $\kappa=0$. It is appropriate to illustrate the behavior of $f(r)$ for some values of parameters. We can adjust the parameters using  relation (\ref{con}) so that the root of the $f(r)$ is located in the region that is more reliable.
For example for the values $Q=1.1, l=1, \kappa=0, q=5$ and $m=1$ the root of the $f(r)$ is $r_h=0.935$ while relation (\ref{con}) shows that the result is more reliable for $r>0.6665$. See Fig. (\ref{fig:ff}). \par 
The Hawking temperature for this black brane is,
  \begin{align}\label{Temp}
  T =\frac{1}{4\pi} \frac{\partial f(r)}{\partial r}\mid_{r=r_h}=\frac{m}{2 \pi  {r_h}^2}+\frac{r_h}{2 \pi l^2}+\frac{q^2}{4 \pi r_h^2}\int^{r_h}u^2  \left(1-e^{\frac{1}{2} W\left(\frac{4 e^{\frac{2 C_1}{q^2}}}{q^2 u^4}\right)}\right)du,
  \end{align}
  This equation can be examined by similar approximations  were considered in obtaining $f(r)$ function. The result is 
  \begin{equation}\label{HT}
	\begin{split}
 		T =\frac{m}{r_{h}^{2}}+\frac{r_{h}}{l^{2}}+\frac{ 2^{\frac{7}{8}}q^{2} r_{h}  \left(\frac{{\mathrm e}^{Q^{2}-1}}{q^{2} r_{h}^{4}}\right)^{\frac{7}{8}} \left(D_1+D_2+D_3+D_4\right)}{240 \pi  \,{\mathrm e}^{Q^{2}-1} \Gamma \! \left(\frac{3}{4}\right)},
 	\end{split}
  \end{equation}
where
 \begin{equation} 
D_1= -5 \,{\mathrm e}^{-\frac{3}{4}} {\mathrm e}^{-\frac{4 Q^{2} q^{2} r_{h}^{4}-7 q^{2} r_{h}^{4}+4 \,{\mathrm e}^{Q^{2}-1}}{4 q^{2} r_{h}^{4}}} \mathrm{WhittakerM}\! \left(\frac{9}{8},\frac{5}{8},\frac{2 \,{\mathrm e}^{Q^{2}-1}}{q^{2} r_{h}^{4}}\right) \Gamma \! \left(\frac{3}{4}\right) q^{4} r_{h}^{8},
 \end{equation}
\begin{equation} 
D_2= 20 \pi  2^{\frac{3}{8}} \left(\frac{{\mathrm e}^{Q^{2}-1}}{q^{2} r_{h}^{4}}\right)^{\frac{1}{8}} \left(\frac{{\mathrm e}^{Q^{2}}}{q^{2} r_{h}^{4}}\right)^{\frac{3}{4}} {\mathrm e}^{-\frac{3}{4}} q^{2} r_{h}^{4},
\end{equation}
 \begin{equation} 
	D_3= -40 \,{\mathrm e}^{-\frac{3}{4}} {\mathrm e}^{Q^{2}-1} {\mathrm e}^{-\frac{4 Q^{2} q^{2} r_{h}^{4}-7 q^{2} r_{h}^{4}+4 \,{\mathrm e}^{Q^{2}-1}}{4 q^{2} r_{h}^{4}}} \mathrm{WhittakerM} \left(\frac{9}{8},\frac{5}{8},\frac{2 \,{\mathrm e}^{Q^{2}-1}}{q^{2} r_{h}^{4}}\right) \Gamma  \left(\frac{3}{4}\right) q^{2} r_{h}^{4},
\end{equation}
and
\begin{align}
D_4&= -64 \,{\mathrm e}^{-\frac{3}{4}} {\mathrm e}^{-\frac{-4 Q^{2} q^{2} r_{h}^{4}+q^{2} r_{h}^{4}+4 \,{\mathrm e}^{Q^{2}-1}}{4 q^{2} r_{h}^{4}}} \mathrm{WhittakerM} \left(\frac{1}{8},\frac{5}{8},\frac{2 \,{\mathrm e}^{Q^{2}-1}}{q^{2} r_{h}^{4}}\right) \Gamma  \left(\frac{3}{4}\right) \nonumber\\&+10 \,2^{\frac{1}{8}} \left(\frac{{\mathrm e}^{Q^{2}-1}}{q^{2} r_{h}^{4}}\right)^{\frac{1}{8}} \Gamma  \left(\frac{3}{4}\right) q^{2} r_{h}^{4}.
\end{align}
  %%%
%Figure (\ref{fig:temp}) shows the approximated behavior of temperature versus horizon  which has been illustrated for the parameter values  $Q=1.1, l=1, q=5$ and $m=1$. 
\begin{figure}[h!]
\centering
\includegraphics[width=6cm]{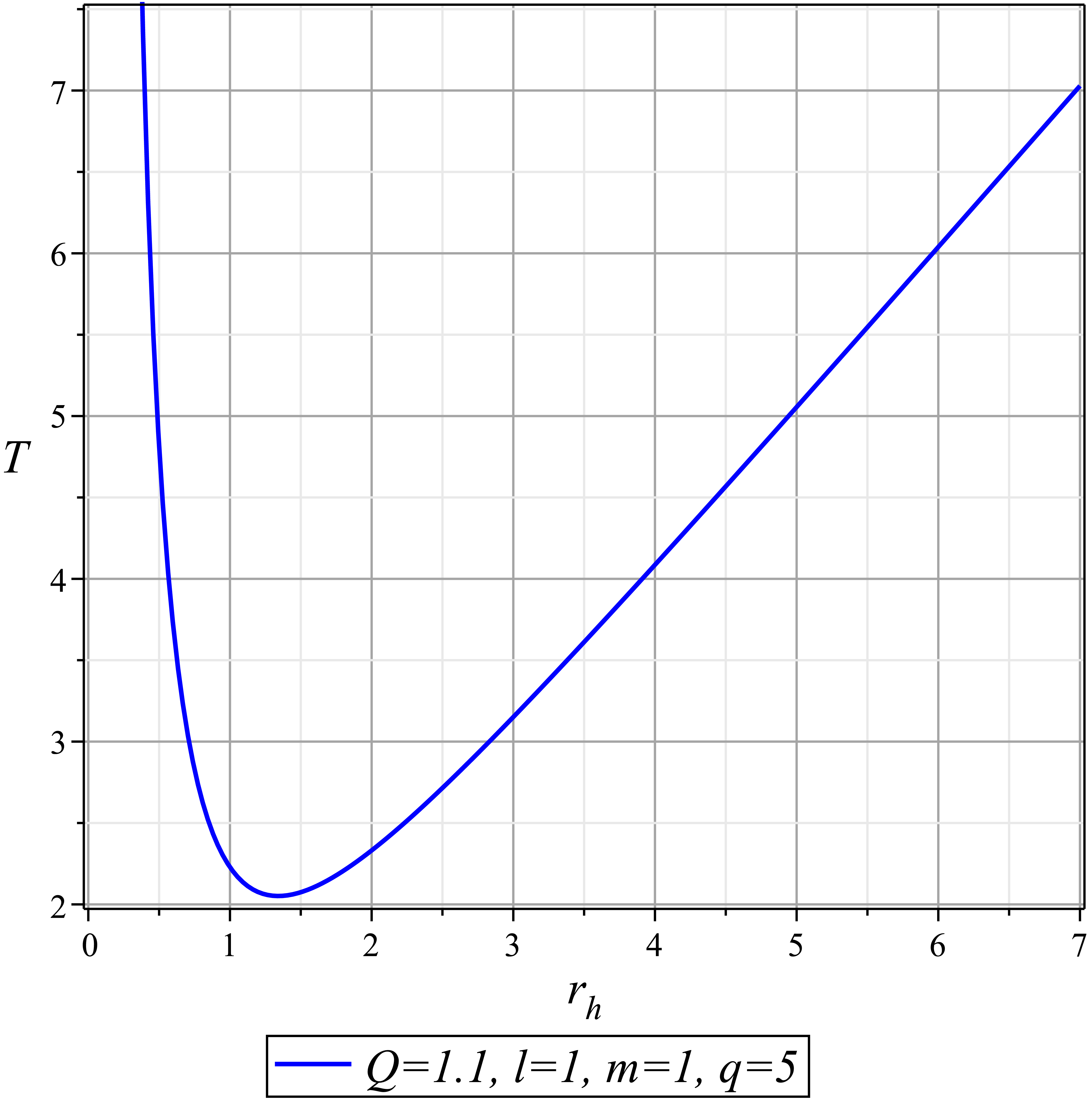}
\caption{ An approximated behavior of temperature for the values $Q=1.1, l=1, q=5$ and $m=1$. It is obvious that a minimum value for the temperature can be seen in this theory. \label{fig:temp}}
\end{figure}
\begin{figure}[h!]
	\centering
	[a]{\includegraphics[width=6cm]{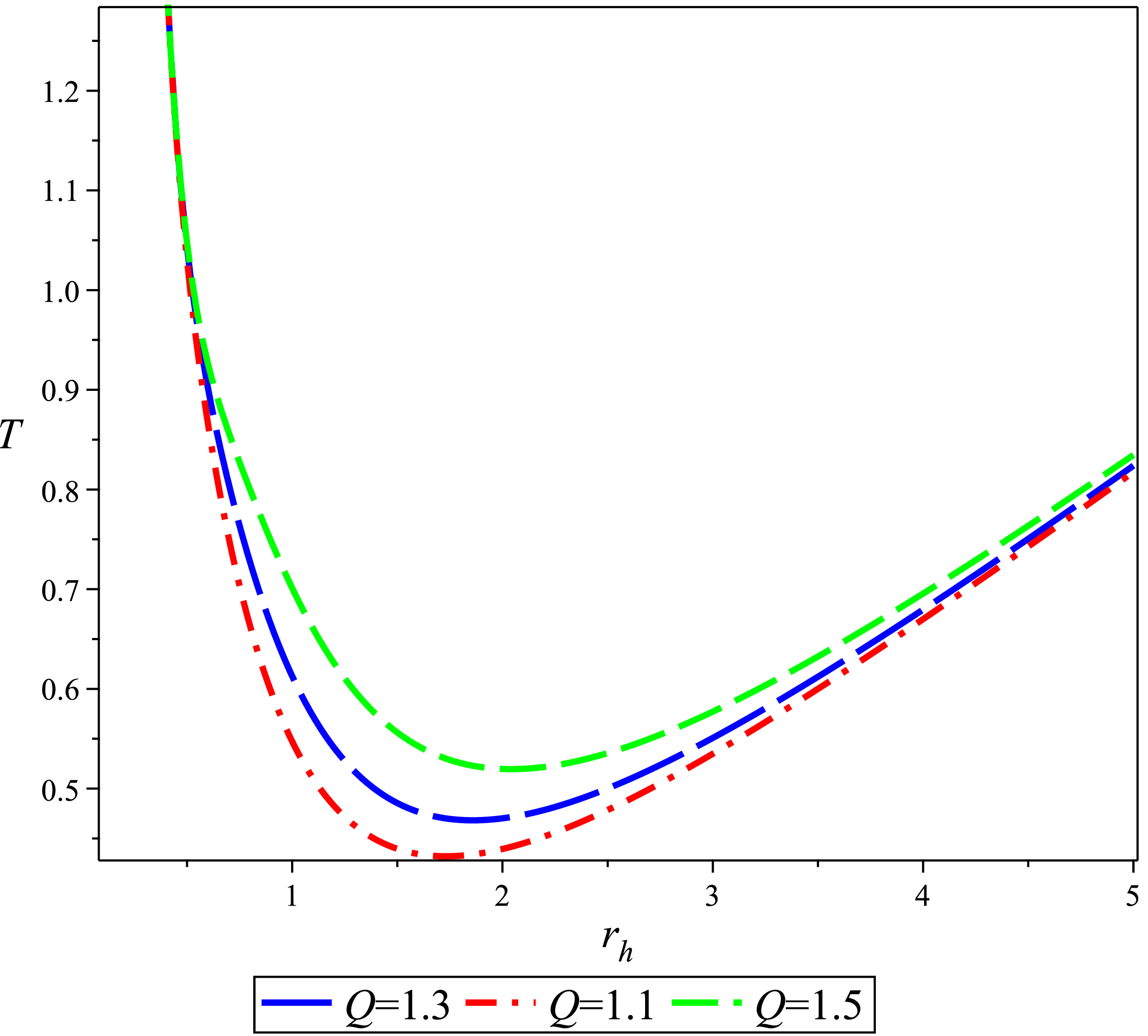}}\quad
	[b]{\includegraphics[width=6.9cm]{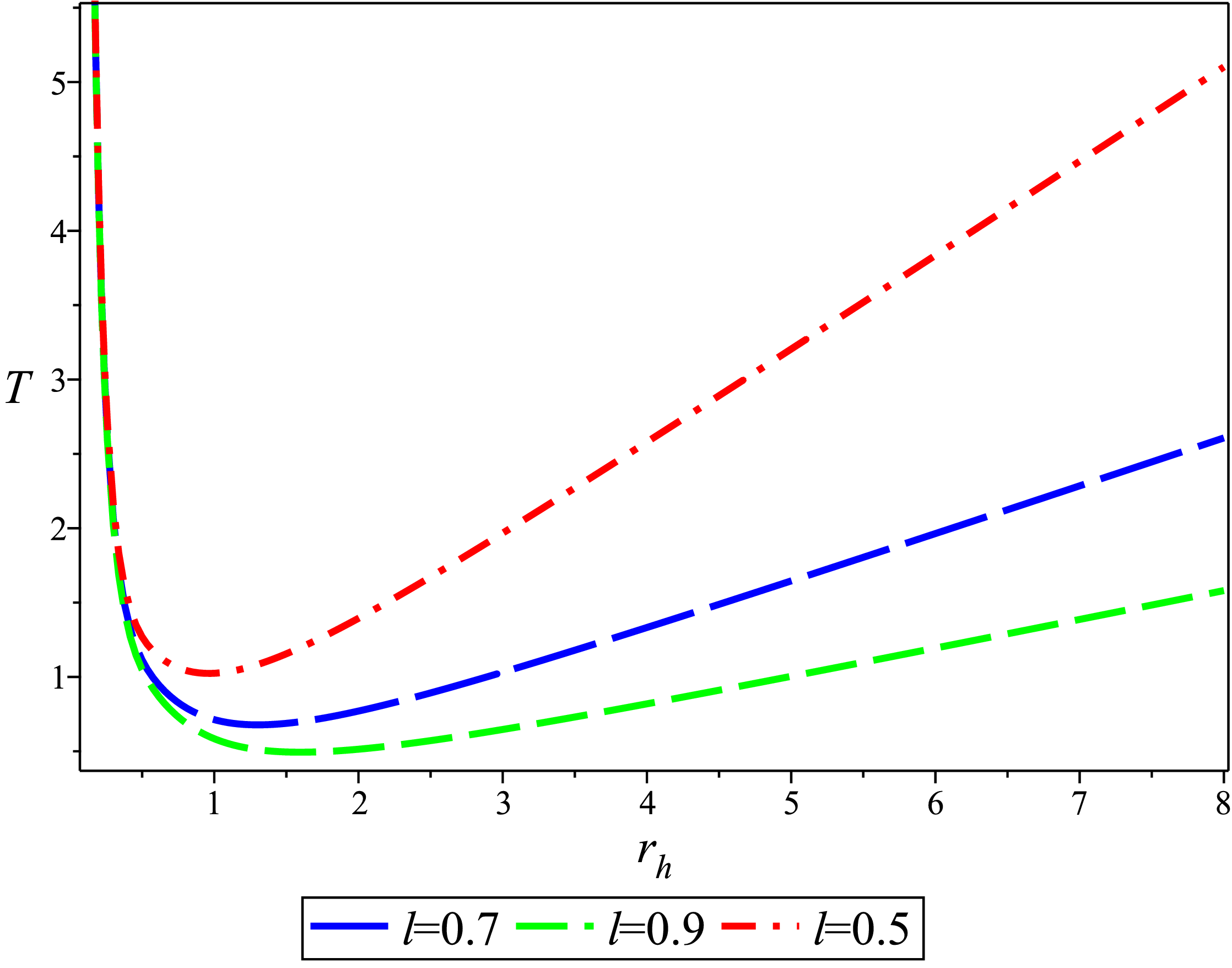}}
	\caption{An approximated behavior of temperature $T$ versus horizon for fixed values $m=1$ and $q=5$ when, \textbf{(a)}: the Yang-Mills charge changes and $l=1$ and \textbf{(b)}: the AdS radius changes and $Q=1.1$. \label{fig:temp}}
\end{figure}
%\begin{figure}[h!]
%\centering
%\includegraphics[width=6cm]{temp.png}
%\caption{ An approximated behavior of temperature for the values $Q=1.1, l=1, q=5$ and $m=1$. It is obvious that a minimum value for the temperature can be seen in this theory. \label{fig:temp}}
%\end{figure}
\begin{figure}[h!]
	\centering
	[a]{\includegraphics[width=6cm]{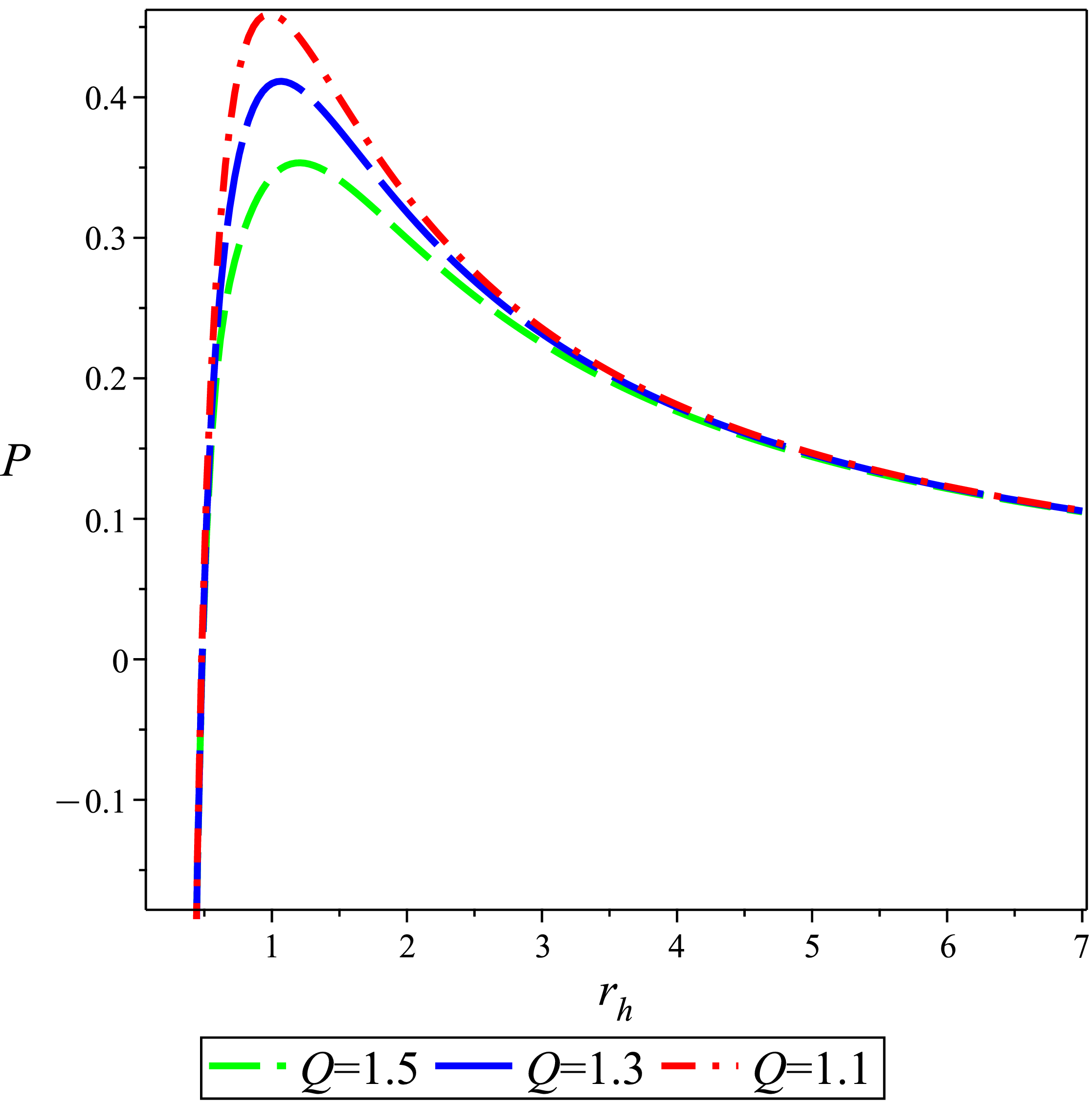}}\quad
	[b]{\includegraphics[width=6cm]{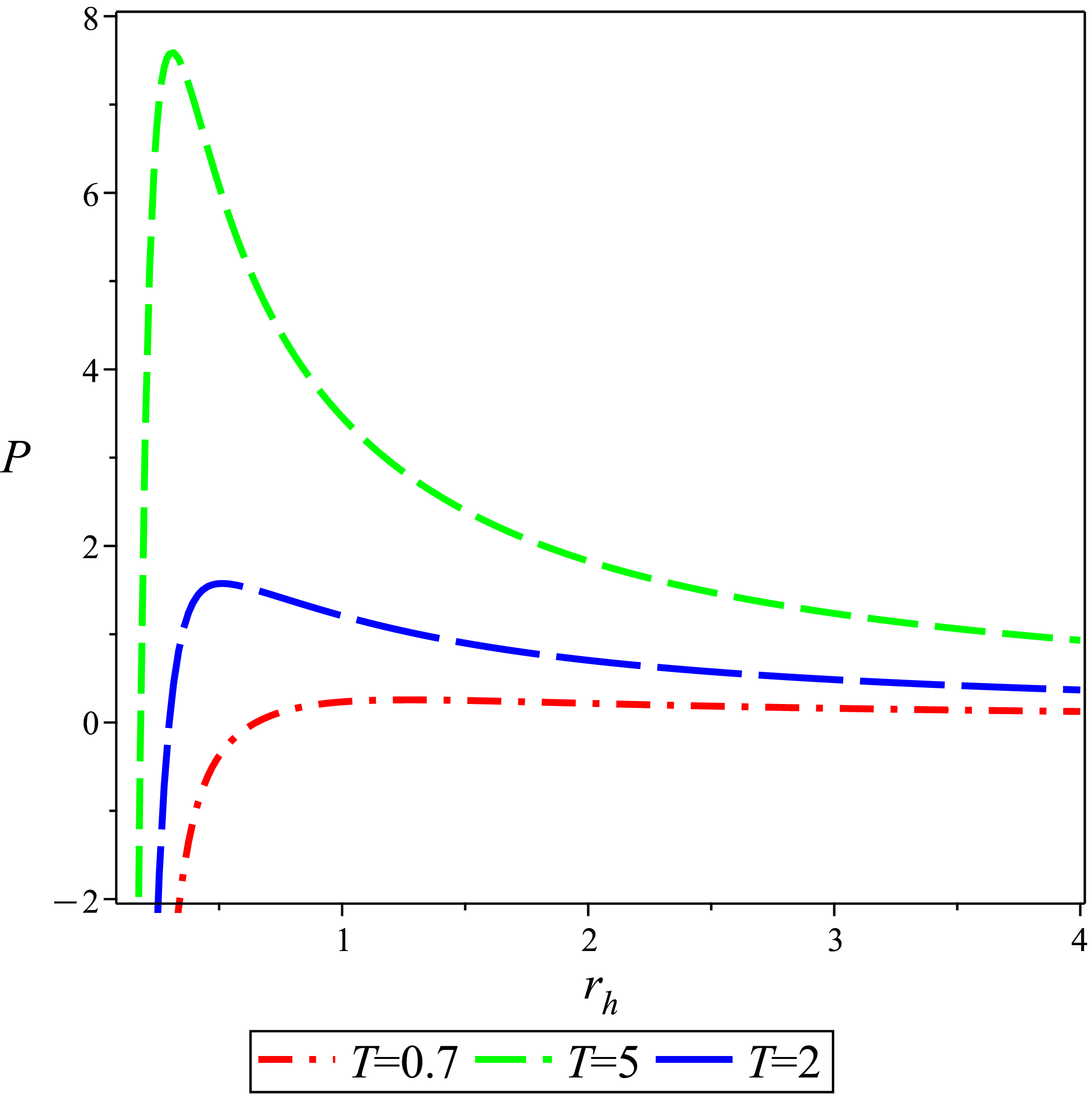}}
	\caption{An approximated behavior of pressure $P$ for fixed values $m=1$ and $q=5$ when, \textbf{(a)}: the Yang-Mills charge changes and $T=1$ and \textbf{(b)}: the temperature changes and $Q=1.1$. \label{fig:pre}}
\end{figure}
Figure (\ref{fig:temp}) shows the approximated behavior of temperature versus horizon.  There is a minimum value for temperature in $T-r_h$ plane like the temperature behavior in the Hawking-Page transition. In the left panel the value of the Yang-Mills charge changes. As the Yang-Mills charge increases, the minimum of the diagram shifts upwards. The right panel shows the temperature behavior when the AdS radius ( proportional to the inverse square of the pressure) changes. As the AdS radius increases, the minimum of the diagrams shifts down. \par 
The behavior of the pressure can be seen in Fig. (\ref{fig:pre}). In the left panel the value of the Yang-Mills charge changes, while in the right panel the temperature of the system varies. The investigations show that there are no critical points in these diagrams like that of the Van der Walls fluids. Nevertheless, it may be a first-order phase transition in this model which can be studied through the investigation of the Gibbs free energy of the system which is not the main subject of this paper.\par 
Usually , to study the thermodynamic stability of the black holes or black branes, the heat capacity of the system is considered.
\begin{figure}[h!]
	\centering
	[a]{\includegraphics[width=6.1cm]{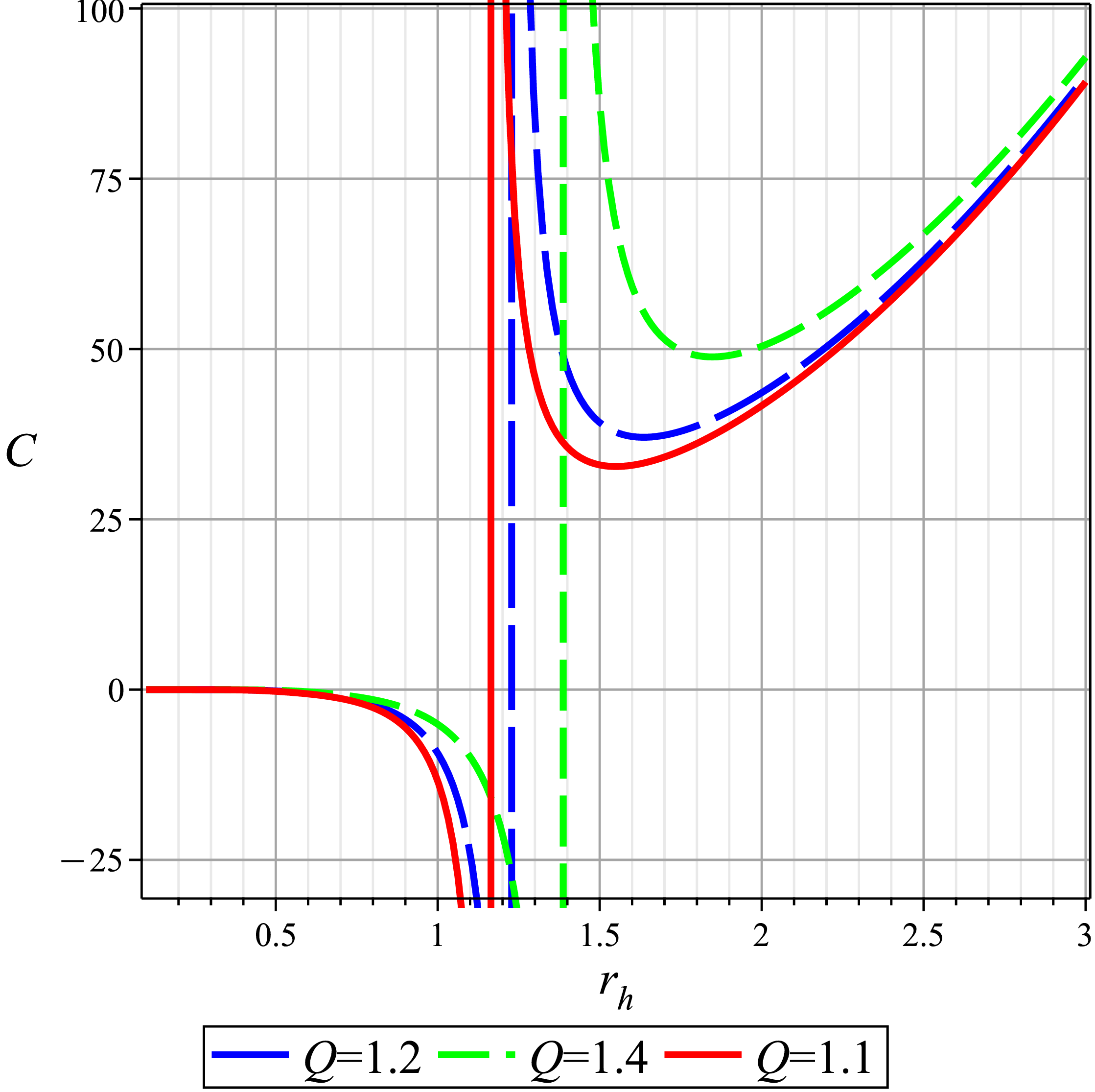}}\quad
	[b]{\includegraphics[width=6cm]{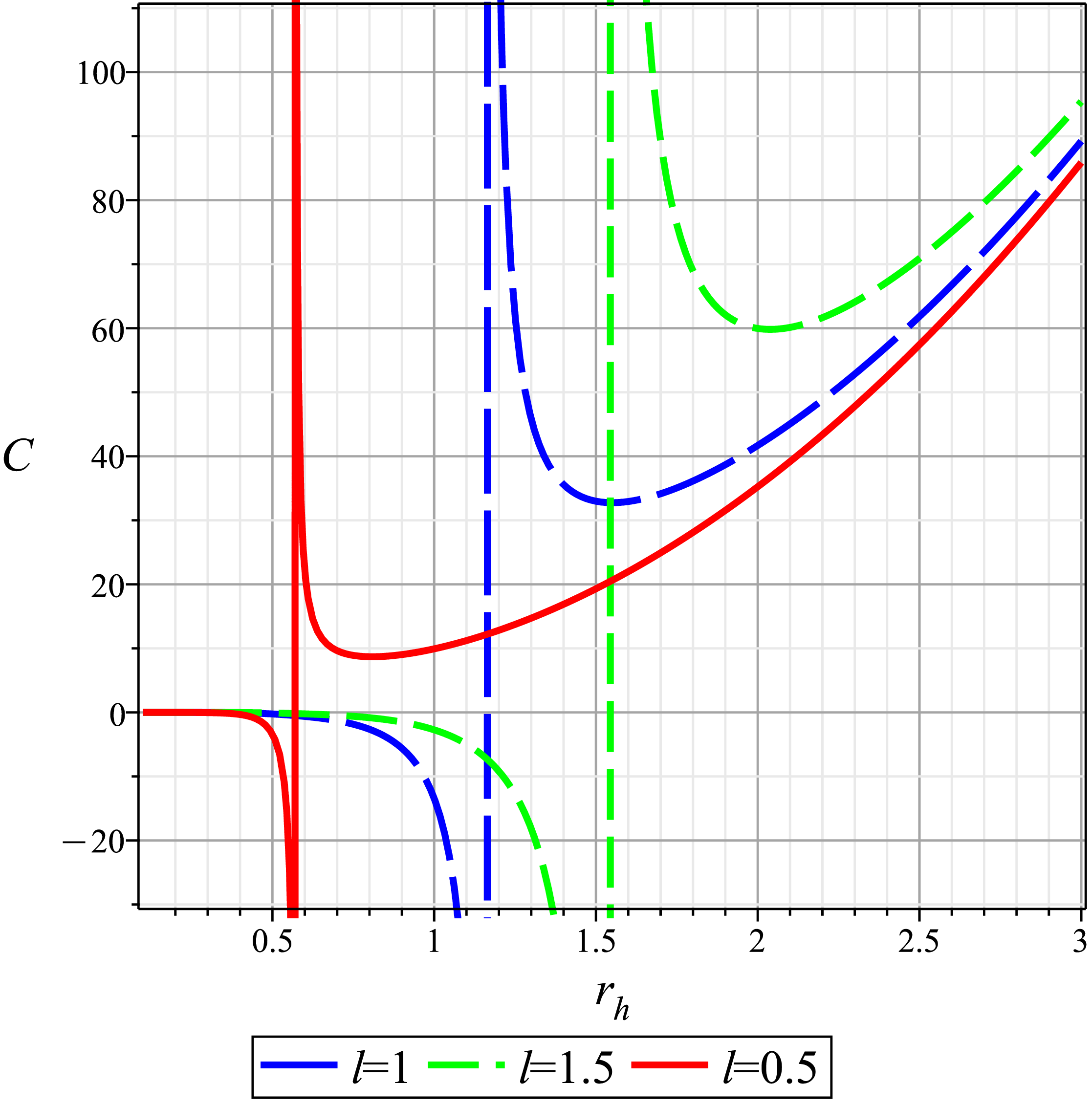}}
	\caption{An approximated behavior of the heat capacity at constant pressure for the fixed values $m=1$ and $q=5$ when, \textbf{(a)}: the Yang-Mills charge changes and $l=1$; \textbf{(b)}: the AdS radius changes and $Q=1.1$. \label{fig:capa}}
\end{figure}
In stable and unstable regions the value of the heat capacity is positive and negative respectively. At the transition point the heat capacity diverges. Our investigations shows the heat capacity diverges only at one point for different values of parameters in this model. After the divergence point the heat capacity is globally positive which shows a global stability. Fig. (\ref{fig:capa}) shows the behavior of the heat capacity. By increasing the value of the Yang-Mills charge and AdS radius, the transition between unstable and stable regions takes place for larger horizons.
\newpage
%%%%%%%%%%%%%%%%%%%%%%%%%%%%%%%%%%%%%%%%%%%%%
\section{Transport Coefficients}
\label{sec3}
According to the AdS/CFT duality, the field $\phi$ in the bulk corresponds to the operator $\mathcal{O}$ on the boundary of the AdS space. Also, the partition function on the gravity side is equivalent to the partition function on the boundary side,
\begin{equation}
Z_{\text{CFT}}=Z_{\text{string}},
\end{equation}
which is realized by GKP–Witten relation\cite{Witten:1998qj},\cite{Gubser:1998bc},\cite{Ferreira-Martins:2019wym} as bellow,
\begin{align}\label{GPKW}
\bigg<\text{exp}{\int_{{\bold{S}}^3} \phi_0 \mathcal{O}}\bigg>_{CFT}=\text{exp}\big(-I_s(\phi)\big),
\end{align}
Here, $\phi_0$ is the value of field on the boundary of AdS and it acts as an external source of a boundary operator $\mathcal{O}$.
We set $\phi=A_{\mu}$ and $\mathcal{O}=J^{\mu}$ in the Eq.(\ref{GPKW}) to calculate DC conductivity.
Therefore, we perturb the gauge field as $A_{\mu} \to A_{\mu}+\tilde{A}_{\mu}$ where the perturbation part of gauge field is chosen as $\tilde{A}_x=\tilde{A}_x(r)e^{-i\omega t}$ \cite{Baggioli:2016oju}. Since, we are in the fluid-gravity duality, it forces us to assume that $\omega$ has small value.
By substituting the perturbed part of the gauge field into  action (\ref{action}) and keeping terms up to the second order of $\tilde{A}$, we will have
\begin{align}\label{action-2}
	S^{(2)}&=\int d^4x \frac{e^{\frac{h'^2}{q^2}}}{ f} \left[-f^2 \left((\partial_r\tilde{A}_x^{(1)})^2+(\partial_r\tilde{A}_x^{(2)})^2+(\partial_r\tilde{A}_x^{(3)})^2\right)\right. \nonumber\\
&\left.-\Big((\tilde{A}_x^{(1)})^2+(\tilde{A}_x^{(2)})^2\Big)(\omega ^2+h^2) + (\tilde{A}_x^{(3)})^2\omega^2\right].
\end{align}
By variation of the action $S^{(2)}$ with respect to $\tilde{A}_x ^a$ we have,
\begin{equation}\label{PerA1}
	q^2 f\left(f \tilde{A}_x^{(1)'}\right)'+q^2(\omega ^2+h^2)\tilde{A}_x^{(1)}+ f^2 \tilde{A}_x^{(1)'} h' h''=0,
\end{equation}
\begin{equation}\label{PerA2}
	q^2 f\left(f \tilde{A}_x^{(2)'}\right)'+q^2(\omega ^2+h^2)\tilde{A}_x^{(2)}+ f^2 \tilde{A}_x^{(2)'} h' h''=0,
\end{equation}
%\begin{equation}
%	f \left(\tilde{A}_x^{(3)'} \left(q^2 f'+2 f h' h''\right)+q^2 f\tilde{A}_x^{(3)''}\right)+q^2 \omega^2 \tilde{A}_x^{(3)}=0    
%\end{equation}
and
\begin{equation}\label{PerA3}
	q^2 f\left(f \tilde{A}_x^{(3)'}\right)'+q^2 \omega^2 \tilde{A}_x^{(3)}+2 f^2 h' h''\tilde{A}_x^{(3)'}=0.    
\end{equation}
First, we solve Eq.(\ref{PerA1}), Eq.(\ref{PerA2}) and Eq.(\ref{PerA3}) near the event horizon. We consider the solutions of $\tilde{A}_x^{(a)}$ as follows,
\begin{align}
	\tilde{A}_x^{(a)}\sim (r-r_h)^{z_a} \, , \qquad a=1,2,3\,\,\,\,,   
\end{align}
where,
\begin{align}\label{z12}
	z_1&=z_2=\pm i \frac{\sqrt{\omega ^2+h(r_h)^2}}{4 \pi T} , \\
	\label{z3}
	z_3&=\pm i \frac{\omega }{4 \pi T}.
\end{align}
%where we use $f\sim4\pi T(r-r_h)$.\\
To solve Eq.(\ref{PerA1}), Eq.(\ref{PerA2}) and Eq.(\ref{PerA3}), we use following ansatzs,
\begin{align}\label{EOMA1}
\tilde{A}_x^{(1)}=\tilde{A}^{(1)}_{\infty}\Big(\frac{-3 f }{\Lambda r^2} \Big)^{z_1}\Big(1+i\omega h_1(r)+\cdots\Big) ,
\end{align}
\begin{align}\label{EOMA2}
\tilde{A}_x^{(2)}=\tilde{A}^{(2)}_{\infty}\Big(\frac{-3 f }{\Lambda r^2}\Big)^{z_2}\Big(1+i\omega h_2(r)+\cdots\Big), 
\end{align}
and
\begin{align}\label{EOMA3}
\tilde{A}_x^{(3)}=\tilde{A}^{(3)}_{\infty}\Big(\frac{-3 f }{\Lambda r^2}\Big)^{z_3}\Big(1+i\omega h_3(r)+\cdots\Big) ,
\end{align}
where $\tilde{A}^{(a)}_{\infty}$ is the value of fields on the boundary and $z_i$'s are the minus branch of (\ref{z12}) and (\ref{z3}).\\
By substituting Eq. (\ref{EOMA1}) into Eq.(\ref{PerA1})  and keeping up to the first order of $\omega$, we obtain,
\begin{align}\label{h1}
&i \pi ^2 q^2 r^2 T^2 h(r_h)^2 h_1((r) f'(r)^2-16 i q^2 r^2 h(r)^2 h_1(r)-\nonumber\\&4 r f \left(\pi  T h_{r_h} h_1(r) \left(q^2 r f''-2 f' \left(q^2-r h' h''\right)\right)+2 q^2 r f'-
\left(\pi  T h(r_h)+2 i\right) h_1'+i \pi ^2 q^2 T^2 h(r_h)^2 h_1 f'\right)-
\nonumber\\&4 f^2 \left(4 i r \left(q^2 r h_1''+h_1' \left(2 r h' h''+i \pi  q^2 T h(r_h)\right)\right)+2 \pi 
T h(r_h) h_1 \left(q^2-2 r h' h''\right)-i \pi ^2 q^2 T^2 h(r_h)^2 h_1\right)=0.
\end{align}
The equation for $h_2(r)$ is the same as $h_1(r)$.\\
By substituting Eq.(\ref{EOMA3}) into Eq.(\ref{PerA3}) and considering the first order of $\omega$, we obtain,
\begin{align}
r \left(f' \left(-2 r h' h''+q^2 \left(r  h_3'-2\right)\right)+q^2 r f''\right)+f \left(-2 r h' h'' \left(r
 h_3'-2\right)+q^2 \left(r^2  h_3''+2\right)\right)=0.
\end{align}
The solution for $h_3(r)$ is as bellow,
\begin{align}
& h_3(r) =C_3 + \int_1^r \Bigg(\frac{2}{u }+\frac{ C_4   e^{\frac {h'(u)^2}{q^2}} - f'(u)}{  f(u)}\Bigg)du,
\end{align}
where $C_3$ and $C_4$ are integration constants. The near horizon behavior of $h_3(r)$  is,
\begin{equation}\label{C4}
h_3 \approx  \bigg(\frac{ C_4   e^{-\frac {h'(r)^2}{q^2}} - 4 \pi T}{  4 \pi T}\bigg) \log(r-r_h)+\text{finite terms}.
\end{equation}
$C_4$ can be determined by demanding regularity of $h_3(r)$ at event horizon. In other words,
\begin{equation}
C_4=4 \pi T {e^{\frac {h'(r)^2}{q^2}}} .
\end{equation}
By substituting the solution of $\tilde{A}_x^{(3)}$ into Eq.(\ref{action-2}) and variation with respect to $\tilde{A}^{(3)}_{\infty}$ , Green's function  can be read as,
\begin{equation} \label{Green1}
 G_{xx}^{(33)} (\omega ,\vec{0})=(\tilde{A}_x^{(3)})^{*}f(r) \partial _{r}\tilde{A}_{x}^{(3)}\bigg|_{r \to \infty}=i \omega C_4.
\end{equation}
By using Kubo formula, $\sigma^{(ab)} _{ij}(k_{\mu})=-\mathop{\lim }\limits_{\omega \to 0} \frac{1}{\omega } \Im G^{(ab)}_{ij}(k_{\mu})$, we have:  
\begin{eqnarray}\label{sigma33}
\sigma_{xx}^{(33)}=-\mathop{\lim }\limits_{\omega \to 0} \frac{1}{\omega } \Im G^{ij}(k_{\mu}) =e^{\frac {h'(r_h)^2}{q^2}}=e^{W[\frac{4 e^{\frac{2 C_1}{q^2}}}{q^2 r_h^4}]}.
\end{eqnarray}
\begin{figure}[h!]
	\centerline{\includegraphics[width=10cm]{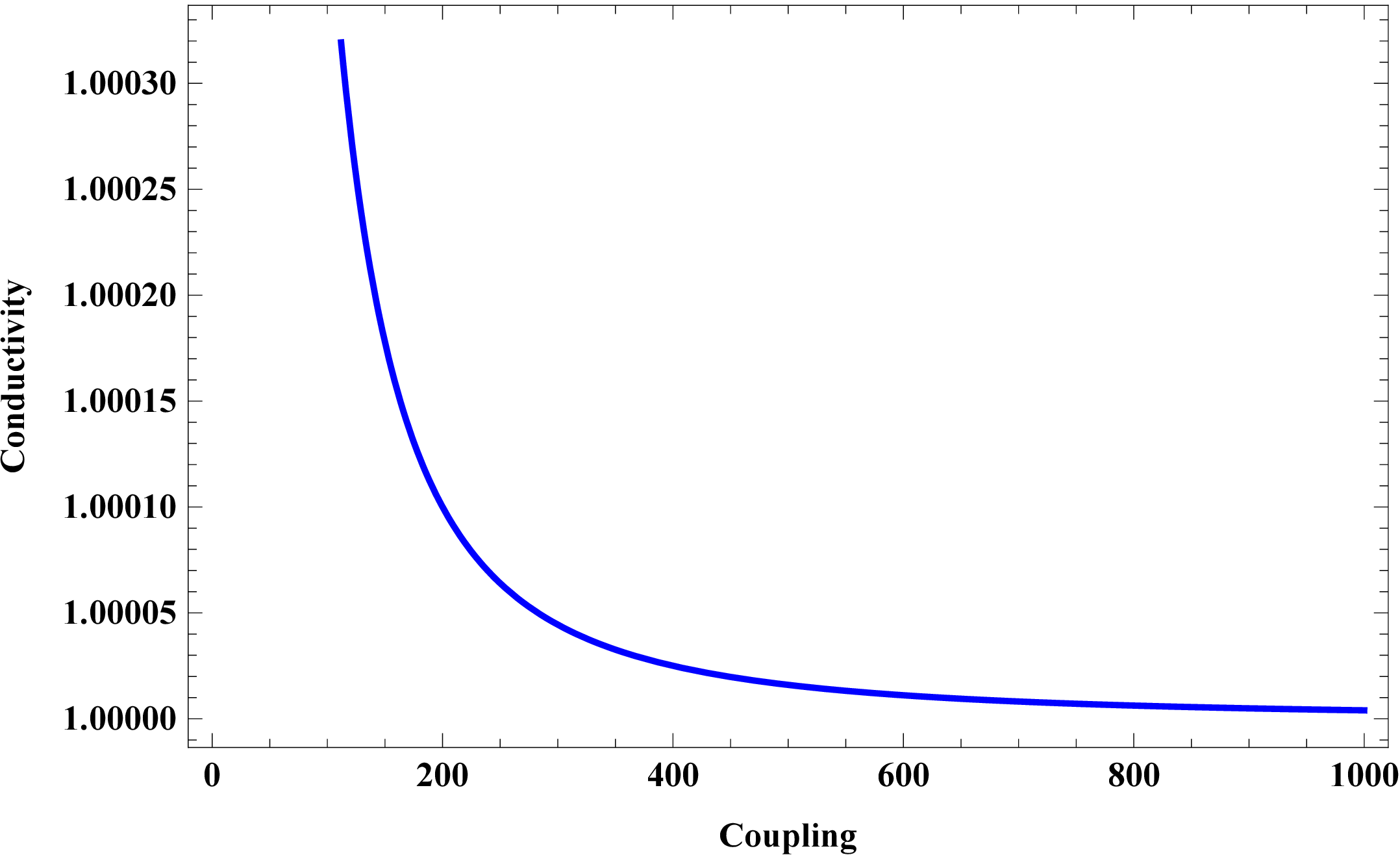}}
	\caption{$\sigma_{xx}^{(33)}-q$ diagram for $r_h=1$. \label{fig:1}}
\end{figure}\\
The conductivity bound is preserved for $ e^{W[\frac{4 e^{\frac{2 C_1}{q^2}}}{q^2 r_h^4}]} > 1 $  and it is violated for\\ $ 0<e^{W[\frac{4 e^{\frac{2 C_1}{q^2}}}{q^2 r_h^4}]} < 1 $.\\\\
In the limit of $q \to \infty$  we have non-abelian Yang-Mills theory and the conductivity bound is saturated.
\begin{eqnarray}
\sigma_{xx}^{(33)} =1.
\end{eqnarray}
The value of $\sigma_{xx}^{(11)}$ and  $\sigma_{xx}^{(22)}$ is calculated by the same procedure of $\sigma_{xx}^{(33)}$ which results in,
\begin{eqnarray}
\sigma_{xx}^{(11)} =\sigma_{xx}^{(22)} =0.
\end{eqnarray}
It means that the color non-abelian DC conductivity in terms of color indices is diagonal and also Ohm's law is diagonal in this model. If we consider the gauge field (\ref{background}) in  $\sigma_{1}$ or $\sigma_{2}$ directions, conductivity will be non-zero in these directions.\\
The ratio of shear viscosity to entropy density is as follows,
\begin{equation}
	\frac{\eta}{s}=\frac{1}{4 \pi} \phi(r_h)^2,
\end{equation}
where $\phi$ is the perturbed part of the metric\cite{Hartnoll:2016tri}.\\
Therefore, we perturb the metric as  $g_{\mu \nu} \to g_{\mu \nu} +\frac{r^2}{l^2} \phi(r )$ \footnote{The components of $t$ and $z$  in  $\phi(t,r,z)$ disappear by Fourier transforms.} and substitute into the action (\ref{action}).  Then, we expand the action up to the second order of $\phi$. Finally, the equation of motion of $\phi(r )$ is given by  varying the resulted action with respect to $\phi(r)$ which leads to,
\begin{eqnarray}
	\phi ' f'+f \phi ''=0.
\end{eqnarray}
The solution of the $\phi(r)$ is as follows,
\begin{align}\label{phi}
	\phi (r)=C_5 + C_6\int^{r}\frac{1 }{ f(u)}du.
\end{align}
We consider the solution of $\phi(r)$ near the event horizon as,
\begin{align}
	\phi (r)=C_5+\frac{C_6}{4 \pi T} \log(r-r_h).
\end{align}
By demanding the regularity of $\phi (r)$ on the event horizon. We have, 
\begin{eqnarray}
	C_6=0.
\end{eqnarray}
By applying the normalization condition on Eq.(\ref{phi}), we have $C_5=1$. So,
\begin{align}
	\phi(r)=1.
\end{align}
Finally, the ratio of shear viscosity to entropy density is given by,
\begin{equation}
	\frac{\eta}{s}=\frac{1}{4 \pi} \phi(r_h)^2=\frac{1}{4 \pi}.
\end{equation}
The ratio $\frac{\eta}{s}$ is satisfied for Einstein-Hilbert gravity as the universal relation $\frac{\eta}{s} \geq \frac{1}{4 \pi}$. This bound which is called KSS bound \cite{Policastro:2002se} can be violated for higher derivative gravitational theories \cite{Sadeghi:2015vaa,Parvizi:2017boc} while, it is saturated for Einstein-Hilbert gravity\cite{Policastro2001}.\\
Our result shows that the KSS bound is saturated for nonlinear exponential Yang-Mills AdS black brane.

%--------------------------------------------------------------------------
 \section{Conclusion}

\noindent We introduced  non-abelian exponential Einstein  Yang-Mills AdS black brane solution and calculated the non-abelian color DC conductivity and the ratio of shear viscosity to entropy density for this model. There is a conjecture that conductivity is bounded by the universal value $\sigma \geq 1$. Our result shows that the conductivity bound is violated for  $ 0<e^{\frac {h'(r_h)^2}{q^2}} < 1 $  and it is preserved for $ e^{\frac {h'(r_h)^2}{q^2}} \ge 1 $ in non-abelian exponential gauge theory but this bound is saturated for Yang-Mills theory\cite{S Parvizi}. The conductivity bound can be violated for a nonlinear model like the exponential model and the violation of conductivity bound is related to Mott insulators.\\
 %depending on the form of the nonlinear theory, or coupling of $F^2$ to the scalar field. Our interpretation is that nonlinearities are somehow similar to electron-electron couplings and therefore the violation related to Mott insulators. It can be concluded that the conductivity is the derivative of the Lagrangian with respect to $F^2$ computed at the event horizon and it is saturated for the Yang-Mills theory but it can be violated for the nonlinear model.\\
The ratio of shear viscosity to entropy density in this model, $\frac{\eta }{s}=\frac{1 }{4 \pi}$, is the same as this value for Einstein-Hilbert gravity. It means that the coupling of the field theory dual to our model  is the same as the coupling of the field theory dual to the Einstein AdS black brane solution, but the color conductivity is different. We note that $\frac{\eta }{s}$ is proportional to the inverse of the square of the field theory coupling i.e. $\frac{\eta }{s} \sim \frac{1 }{\lambda^2}$.
%The ratio of shear viscosity to entropy density for this model is $\frac{\eta }{s}=\frac{1 }{4 \pi}$. Therefore, the Kovtun-Son-Starinet (KSS) bound \cite{Kovtun:2004de} that it stated this value for Einstein-Hilbert gravity is $\frac{\eta }{s}=\frac{1 }{4 \pi}$ is preserved for this model. $\frac{\eta }{s}$ is proportional to inverse squared coupling of field theory, $\frac{\eta }{s} \sim \frac{1 }{\lambda^2}$. It means that the coupling of the field theory dual to our model  is the same as the coupling of the field theory dual to the Einstein AdS black brane solution, but the color conductivity is different.

%-------------------------------------------------------------------------
\vspace{1cm}
\noindent {\large {\bf Acknowledgment} } We would like to thank Ahmad Moradpouri for useful comments and suggestions.

%-------------------------------------------------------------------------
\vspace{1cm}
\noindent {\large {\bf Data Availability statement } } \\\\
 All data that support the findings of this study are included within the article (and any supplementary
files).
%-------------------------------------------------------------------------
%\vspace{1cm}
%\noindent {\large {\bf Competing interests }  The author declares there are no competing interests.
%-------------------------------------------------------------------------

\end{document}